\documentclass[11pt,a4paper]{article}
\pdfoutput=1
\usepackage{jheppub}

\usepackage{amsmath}
\usepackage{verbatim}
\usepackage{amssymb}
\usepackage{inputenc, array}
\usepackage{textcomp}
\usepackage{appendix}
\usepackage{mathrsfs}
\usepackage[dvipsnames]{xcolor}
\newcommand{\e}{\epsilon}
\newcommand{\be}[1]{\begin{equation}\label{#1} }
\newcommand{\ee}{\end{equation}}
\newcommand{\bea}[1]{\begin{eqnarray}\label{#1} }
\newcommand{\eea}{\end{eqnarray}}
\newcommand{\bes}[1]{\begin{subequations}\label{#1} }
\newcommand{\ees}{\end{subequations} }
\newcommand{\p}{\partial}
\newcommand{\refb}[1]{(\ref{#1})}

\renewcommand{\L}{{\bar{L}}}

\newcommand{\z}{{\bar z}}
\newcommand{\h}{{\bar h}}

\renewcommand{\>}{\rangle}
\newcommand{\<}{\langle}

\renewcommand{\(}{\left(}
\renewcommand{\)}{\right)}

\newcommand{\D}{\Delta}

\renewcommand{\(}{\left(}
\renewcommand{\)}{\right)}

\title{Scattering Amplitudes: Celestial and Carrollian}

\author[a]{Arjun Bagchi,} \author[b,c]{Shamik Banerjee,}  \author[d]{Rudranil Basu,} \author[a]{and Sudipta Dutta.}
\author{\\}

\affiliation[a]{Indian Institute of Technology Kanpur, Kanpur 208016, INDIA.\\} 

\affiliation[b]{Institute of Physics, Sachivalaya Marg, Bhubaneshwar 751005, INDIA.\\} 

\affiliation[c]{Homi Bhabha National Institute, Anushakti Nagar, Mumbai 400085, INDIA.\\}

\affiliation[d]{Department of Physics, BITS-Pilani, K K Birla Goa Campus, Zuarinagar, Goa-403726.
	INDIA.\\}

\emailAdd{abagchi@iitk.ac.in, banerjeeshamik.phy@gmail.com, rudranilb@goa.bits-pilani.ac.in, dsudipta@iitk.ac.in}

\preprint{}

\abstract{Recent attempts at the construction of holography for asymptotically flat spacetimes have taken two different routes. Celestial holography, involving a two dimensional (2d) CFT dual to 4d Minkowski spacetime, has generated novel results in asymptotic symmetry and scattering amplitudes. A different formulation, using Carrollian CFTs, has been principally used to provide some evidence for flat holography in lower dimensions. Understanding of flatspace scattering has been lacking in the Carroll framework. In this work, using ideas from Celestial holography, we show that 3d Carrollian CFTs living on the null boundary of 4d flatspace can potentially compute bulk scattering amplitudes. 3d Carrollian conformal correlators have two different branches, one depending on the null time direction and one independent of it. We propose that it is the time-dependent branch that is related to bulk scattering. We construct an explicit field theoretic example of a free massless Carrollian scalar that realises some desired properties.}

\begin{document}
\maketitle

\section{Introduction}
The Holographic Principle has been one of our primary routes to a theory of quantum gravity, formulated in terms of a lower dimensional field theory. Although there has been a great deal of success in understanding holography in Anti de Sitter spacetime through AdS/CFT, a similar understanding for the apparently more straight-forward case of asymptotically flat spacetimes is lacking. There is, however, a consorted recent effort at rectifying this situation. There are two principal avenues of addressing this problem - Celestial Holography and Carrollian Holography. Bondi-Metzner-Sachs (BMS) \cite{Bondi:1962px} symmetries, which arise as asymptotic symmetries of flat spacetimes based on the null boundary, are important to both approaches. 

\medskip

Celestial holography has grown out of the basic observation due to Strominger \cite{Strominger:2013jfa} that in asymptotically flatspace time soft theorems for $S$-matrix elements can be thought of as Ward identities for asymptotic symmetries \cite{He:2014laa,Strominger:2014pwa,Kapec:2014opa,Kapec:2016jld,He:2017fsb,Ball:2019atb,Kapec:2017gsg,Fotopoulos:2019vac,Stieberger:2018onx}. This correspondence is holographic in nature. The fundamental claim of celestial holography is that there is a two dimensional (2d) CFT on the celestial sphere which computes the scattering amplitudes for processes taking place in the four dimensional asymptotically flat space time. This computation is facilitated by writing the $S$-matrix in boost eigenstates \cite{deBoer:2003vf,Pasterski:2016qvg,Pasterski:2017kqt,Banerjee:2018gce,Banerjee:2018fgd} in which the Ward identities for asymptotic symmetries take the well known form of Ward identities in a 2d CFT. This CFT is known as the Celestial CFT \footnote{Please see Appendix \ref{cel} for a brief review of celestial amplitudes.}. This approach to the flat space holography has already produced many novel results about asymptotic symmetries and scattering amplitudes \cite{Banerjee:2020zlg,Banerjee:2020vnt,Guevara:2021abz,Strominger:2021lvk,Himwich:2021dau,Pasterski:2022joy,Jiang:2021csc,Campiglia:2014yka,Banerjee:2021dlm,Donnay:2020guq,Freidel:2021ytz,Freidel:2021dfs} in four dimensions. The reader is pointed to the excellent recent reviews \cite{Strominger:2017zoo, Pasterski:2021rjz,Raclariu:2021zjz} for more details on Celestial holography.


\medskip

Another school of thought has been the attempt to build duals of asymptotitcally flat spacetime in terms of a one-dimensional lower field theory that enjoys BMS symmetry. These field theories are conformal theories living on the null boundary of spacetime and can be understood as Carroll contractions of usual relativistic CFTs, which takes the speed of light $c$ to zero \cite{Bagchi:2010zz, Bagchi:2012cy}. We shall call this approach Carroll holography. The success of this formulation has principally been in the three dimensional bulk and two dimensional field theories, where various checks have been performed between the boundary and the bulk, including the matching of entropy \cite{Bagchi:2012xr,Barnich:2012xq,Bagchi:2013qva}, stress-tensor correlations \cite{Bagchi:2015wna}, entanglement entropy \cite{Bagchi:2014iea,Jiang:2017ecm,Hijano:2017eii}. Some other important advances are \cite{Barnich:2012aw, Bagchi:2012yk, Duval:2014lpa, Hartong:2015usd, Bagchi:2016geg, Bagchi:2021qfe} and higher dimensional explorations include \cite{Bagchi:2016bcd, Ciambelli:2018wre, Bagchi:2019xfx}.
Crucially, the understanding of scattering processes has been lacking in this formulation. 

\medskip

In this paper, we will provide a bridge between the two formulations. We will show that using BMS or Conformal Carroll symmetries in a 3d field theory living on null infinity, one can formulate the scattering problem in 4d asymptotically flat spacetimes. We will further demonstrate the plausibility of our proposal by constructing an explicit realisation of Carrollian CFTs in terms of a 3d massless Carroll scalar with some desired features. 

\paragraph{Note added:} When this paper was being readied for submission, \cite{Donnay:2022aba} appeared on the arXiv. Although both papers attempt to link Carroll and Celestial holography, our approaches are complementary. 

\section{BMS and Carroll CFTs}
As is now well known, and has been known since the 1960s, the symmetries of interest in asymptotically flat spacetimes in $d=4$ actually extends beyond the Poincare group to an infinite dimensional group discovered initially by Bondi, van der Burgh, Metzner and Sachs \cite{Bondi:1962px}. The BMS symmetry algebra of 4d flat spacetime at its null boundary $\mathscr{I}^\pm$ is given by:
\bes{bms4}
\begin{align}
 [L_n,L_m]&=(n-m)L_{n+m} ,\quad [\bar{L}_n,\bar{L}_m]=(n-m)\bar{L}_{n+m} \\ 
 [L_n,M_{r,s}]&=\(\frac{n+1}{2}-r\)M_{n+r,s} , \quad [\bar{L}_n,M_{r,s}]=\(\frac{n+1}{2}-s\)M_{r,n+s} \\ 
 [M_{r,s},M_{t,u}]&=0.
\end{align}
\ees
Here $M_{r,s}$ are the generators of infinite dimensional angle dependent translations at $\mathscr{I}^\pm$ known as supertranslations. The original BMS group was given by these infinite dimensional supertranslations on top of the usual Lorentz group denoted here by the generators $\{L_0, L_{\pm 1}, \bar{L}_0, \bar{L}_{\pm 1} \}$. Following \cite{Barnich:2009se, Barnich:2010eb}, there has been an effort to consider the full conformal group on the sphere at infinity and hence all modes of the $L_n$ generators, the so-called super-rotations{\footnote{There exists other extensions e.g.\cite{Campiglia:2014yka, Compere:2018ylh}.}}.

\medskip

In 2d Celestial CFT superrotation or local conformal transformations on the celestial sphere are generated by a stress tensor which is the shadow transform of the subleading soft graviton \cite{Cachazo:2014fwa,Kapec:2014opa,Fotopoulos:2019vac}. After shadow transformation the subleading soft graviton theorem \cite{Cachazo:2014fwa} becomes the well known Ward identity for stress tensor in a 2d CFT. 

\medskip

Let us now discuss 3d Carrollian CFT. We are interested in defining a 3d conformal field theory on $\mathscr{I}^+$ which is topologically $\mathbb{R}_u \times \mathbb{S}^2$, where $\mathbb{R}_u$ is a null line and $\mathbb{S}^2$ is the sphere at infinity.  The null line makes the induced metric of $\mathscr{I}^+$  degenerate. Hence the Riemannian structure is replaced by a so-called Carrollian structures on the intrinsic geometries of these hypersurfaces. CFTs living on $\mathscr{I}^\pm$ are naturally expected to be invariant under the conformal isometries of these Carrollian structures. We refer the reader to Appendix \ref{car} for more details on Carrollian and conformal Carrollian isometries. Rather intriguingly, conformal Carrollian symmetries have been shown to be isomorphic to BMS symmetries in one higher dimension \cite{Duval:2014lpa, Bagchi:2010zz} 
\be{}
\mathfrak{CCarr}_d = \mathfrak{bms}_{d+1}.
\ee
Hence a 3d Carrollian CFT naturally realises the extended infinite-dimensional BMS$_4$ symmetry. These 3d Carrollian CFTs would be our field theories of interest, which we will show to be a potential candidate for a holographic description of scattering amplitudes in 4d asymptotically flat spacetimes.

\medskip

For these 3d theories, a particular useful representation of vector fields to consider is \cite{Bagchi:2016bcd}:
\begin{align}
L_n = -z^{n+1}\p_z - \frac{1}{2}(n+1)z^n u \p_u \quad  \L_n = -\z^{n+1}\p_z - \frac{1}{2}(n+1)\z^n u \p_u \quad M_{r,s} = z^r \z^s \p_u 
\end{align}
Here $z, \z$ are stereographic coordinates on the sphere. We will label the Carroll conformal fields $\Phi$ living on $\mathscr{I}^+$ with their weights under $L_0$ and $\L_0$: 
\be{}
[L_0, \Phi(0)] = h \Phi(0), \quad [\L_0, \Phi(0)] = \h \Phi(0). 
\ee
We will assume the existence of Carrollian primary fields living on $\mathscr{I}^+$. The primary conditions are  \cite{Bagchi:2016bcd, Banerjee:2020kaa}:
\be{}
[L_n, \Phi(0)] = 0, \quad [\L_n, \Phi(0)] = 0, \quad \forall n>0, \quad [M_{r,s}, \Phi(0)] = 0, \quad \forall r, s>0.
\ee
In particular, it is important to stress that the last condition is an additional requirement on these fields, which is unlike a 2d CFT. Also for the supertranslations, any one of $r$ or $s$ being greater than zero annihilates the primary field. The transformation rules of the {\em three dimensional} Carrollian primary fields $\Phi_{h, \h}(u, z, \z)$ at an arbitrary point on $\mathscr{I}^+$ under the infinitesimal BMS transformations are given by
\bes{trans}
\begin{align}
\delta_{L_n} \Phi_{h, \h}(u, z, \z) &= \e \left[z^{n+1}\p_z + (n+1) z^n \left(h+ \frac{1}{2} u\p_u\right) \right] \Phi_{h, \h}(u, z, \z), \\
\delta_{M_{r,s}} \Phi_{h, \h}(u, z, \z) &= \e z^{r}\z^s \p_u \Phi_{h, \h}(u, z, \z).
\end{align}
\ees
There is a similar relation for the antiholomorphic piece. 

\medskip

Let us now discuss how the structure of a Carrolian CFT that we have discussed above fits into the framework of Celestial Holography.

\section{Relation to 4d scattering amplitudes via Celestial Holography}
As we have stressed above, one of the main reasons for studying Carrollian CFTs is that its symmetries are the same as the extended BMS algebra. So potentially Carroll CFTs can be a holographic dual of the quantum theory of gravity in asymptotically flat spacetime. Now we know, from general considerations, that the only observables in a quantum theory of gravity in asymptotically flat space time are the $S$-matrix elements. Therefore, given a holographic dual, one should be able to compute the spacetime $S$-matrix from this. Moreover, if the dual is a field theory or at least looks like one then presumably the $S$-matrix elements should be somehow related to the correlation functions of the field theory. This is the point of view that we adopt in this paper. 

\medskip

In the next section, we will focus on the correlation functions of the Carrollian CFTs. We will find that there are two kinds of correlation functions or two branches. In one branch, the correlation functions are independent of the null time direction{\footnote{The null direction of Carrollian field theories can be interpreted as the time direction. One of the reasons for this is that the Carroll limit from the relativistic theories involves a contraction of the time direction.}}  and have the structure of correlation functions of a 2d CFT. However, in the other branch the correlators have explicit (null) time dependence and do not look like those of a 2d CFT. For example, unlike 2d CFT, the two point function  in this branch is {\em{ultra-local in the spatial directions}} and nonzero even when the scaling dimensions of the operators are different. Similarly one can show using the 4d Poincare or global Conformal Carroll invariance of the Carrollian CFT that the time dependent three point function is zero. This problem can be solved if we treat $z$ and $\bar z$ as independent complex coordinates rather than complex conjugates of each other. These are reminiscent of the properties of scattering amplitudes of massless particles in 4d flat spacetime. So what is the relation of Carroll CFT correlations to scattering amplitudes? In this paper we propose an answer. 

\medskip

In order to answer this question, we use ideas from Celestial holography. (For a quick recap of the essential features of Celestial holography, the reader is pointed to Appendix~\ref{cel}). In Celestial holography the dual theory is conjectured to be a 2d (relativistic) CFT which lives on the celestial sphere. The important point for our purpose is that the correlation functions of the celestial CFT are given by the Mellin transform of the 4d scattering amplitudes \cite{deBoer:2003vf,Pasterski:2016qvg,Pasterski:2017kqt,Banerjee:2018gce,Banerjee:2018fgd}. Let us briefly describe this. For simplicity let us consider only massless particles. 

We parametrize the four momentum of a massless particle as
\be{pmu}
p^{\mu} = \omega \( 1+ z\bar z, z+\bar z, -i(z-\bar z), 1- z\bar z\), \  p^\mu p_{\mu} = 0
\ee
We also introduce a symbol $\epsilon$ which is equal to $\pm 1$ if the particle is (outgoing) incoming. 

Using this parametrization, the Mellin transformation can be written as \cite{Pasterski:2016qvg,Pasterski:2017kqt}, 
\be{mel}
\mathcal{M}\(\{ z_i, \bar z_i, h_i, \bar h_i, \epsilon_i \}\) = \prod_{i=1}^n \int_0^{\infty} d\omega_i \omega_i^{\D_i -1} S\(\{\epsilon_i\omega_i,z_i,\bar z_i, \sigma_i\}\), \  \D\in \mathbb{C}, \  \sigma\in\frac{\mathbb Z}{2}
\ee
where $S$ is the $S$-matrix element for $n$ massless particle scattering. Here we have also defined
\be{}
h=\frac{\D +\sigma}{2}, \  \bar h = \frac{\D-\sigma}{2}
\ee

One can show \cite{Pasterski:2016qvg,Pasterski:2017kqt} using the Lorentz transformation property of $S$-matrix that the object $\mathcal{M}$ on the LHS indeed transforms like the correlation function of $n$ primary operators of weight $(h,\bar h)$ in a 2d CFT \footnote{Note that $\sigma$ is the 4d helicity and also the 2d spin of the corresponding operator.}. After Mellin transformation the coordinates $(z,\bar z)$ can be interpreted as the stereographic coordinates of the celestial sphere and physically represent the direction of motion of the massless particle. For our purpose however, we will use a modification \cite{Banerjee:2018gce,Banerjee:2018fgd} of \eqref{mel} such that the correlation function $\mathcal{M}$ is now defined on a 3d space with coordinates $(u,z,\bar z)$. This space can be interpreted as the (future) null-infinity with $u$ as the retarded time and $(z,\bar z)$ as the stereographic coordinates of the celestial sphere. One can show \cite{Banerjee:2018gce,Banerjee:2020kaa,Banerjee:2020zlg,Banerjee:2018fgd} that under supertranslation, 
\be{}
u \rightarrow u' = u + f(z,\bar z), \  z \rightarrow z' = z, \  \bar z \rightarrow \bar z' = \bar z
\ee
Similarly under superrotation or local conformal transformations 
\be{prime}
u \rightarrow u' = \(\frac{dw}{dz}\)^{\frac{1}{2}}  \(\frac{d\bar w}{d\bar z}\)^{\frac{1}{2}} u, \  z\rightarrow z' = w(z), \  \bar z \rightarrow \bar z' = \bar w(\bar z)
\ee
Now the modified transformation \cite{Banerjee:2018gce,Banerjee:2018fgd} has the following form:
\be{modmellin}
\mathcal{\tilde {M}}\(\{ u_i, z_i, \bar z_i, h_i, \bar h_i, \epsilon_i \}\) = \prod_{i=1}^n \int_0^{\infty} d\omega_i \omega_i^{\D_i -1} e^{-i \epsilon_i \omega_i u_i} S\(\{\epsilon_i\omega_i,z_i,\bar z_i, \sigma_i\}\), \  \D\in \mathbb{C}
\ee
One can show \cite{Banerjee:2018gce,Banerjee:2020kaa,Banerjee:2020zlg,Banerjee:2018fgd} using the celebrated Soft Theorem - Ward Identity correspondence \cite{Strominger:2013jfa,He:2014laa,Strominger:2014pwa,Kapec:2014opa,Kapec:2016jld,He:2017fsb,Ball:2019atb,Kapec:2017gsg,Fotopoulos:2019vac} that $\tilde{\mathcal M}$ transforms covariantly under the extended BMS$_4$ transformations. In Celestial holography the modified Mellin transformation \eqref{modmellin} is used to compute the graviton celestial amplitudes in general relativity because the original Mellin transformation integral \eqref{mel} is not convergent due to bad UV behaviour of graviton scattering amplitudes in GR. It turns out that instead when \eqref{modmellin} is used the time coordinate $u$ acts as a UV regulator and as a result $\mathcal{\tilde M}$ is finite. For more details the reader is referred to \cite{Banerjee:2020zlg,Banerjee:2020kaa,Banerjee:2019prz}.

\medskip

Now it is useful to write the modified celestial amplitude $\tilde{\mathcal M}$ as a correlation function of fields defined on null infinity. So following \cite{Banerjee:2018gce} we define
\be{confprim1}
\phi^{\epsilon}_{h,\bar h}(u,z,\bar z) = \int_{0}^{\infty} d\omega \  \omega^{\D-1} e^{-i\epsilon \omega u} a(\epsilon\omega, z, \bar z, \sigma).
\ee
where $a(\epsilon\omega, z,\bar z, \sigma)$ is the momentum space (creation) annihilation operator of a massless particle with helicity $\sigma$ when $\(\epsilon = -1\)$ $\epsilon = 1$. In terms of these fields we can write
\be{}
\mathcal{\tilde {M}}\(\{ u_i, z_i, \bar z_i, h_i, \bar h_i, \epsilon_i \}\) = \< \prod_{i=1}^n \phi^{\epsilon_i}_{h_i,\bar h_i}(u_i,z_i,\bar z_i)\>.
\ee
Now, the field $\phi^{\epsilon}_{h,\bar h}(u,z,\bar z)$ transforms covariantly under the extended BMS$_4$ transformation. Under superrotation \cite{Banerjee:2018gce,Banerjee:2020kaa,Banerjee:2020zlg,Banerjee:2018fgd}
\be{vir}
\phi^{\epsilon}_{h,\bar h}(u,z,\bar z) \rightarrow \(\frac{dw}{dz}\)^h \(\frac{d\bar w}{d\bar z}\)^{\bar h} \phi^{\epsilon}_{h,\bar h}(u',z',\bar z')
\ee
where the primed coordinates are defined in \eqref{prime}. Similarly under supertranslation,
\be{trans2}
\phi^{\epsilon}_{h,\bar h}(u,z,\bar z) \rightarrow \phi^{\epsilon}_{h,\bar h}(u + f(z,\bar z),z,\bar z)
\ee
It is easy to see that for infinitesimal BMS$_4$ transformations \eqref{vir} and \eqref{trans2} reduce to the equations \eqref{trans} written in terms of the primaries of a \textit{Carrollian} CFT. 

\medskip

Therefore, it is not unreasonable to wonder whether one can identify the Carrollian primaries with the primaries $\phi^{\epsilon}_{h,\bar h}(u,z,\bar z)$ of Celestial Holography. If this is true then this will open the road towards connecting the Carrollian CFT correlation functions with bulk scattering amplitudes because the field $\phi^{\epsilon}_{h,\bar h}(u,z,\bar z)$ is directly related to standard creation-annihilation operators by \eqref{confprim1}. 

\section{The Proposal}

\subsection{The central claim}

Our central claim in this paper is the following: \\

{\textit{It is natural to identify the time-dependent correlation functions of primaries in a Carrollian CFT with the modified Mellin amplitude}} $$\mathcal{\tilde M}\(\{ u_i, z_i, \bar z_i, h_i, \bar h_i, \epsilon_i \}\) = \prod_{i}\<\phi^{\epsilon_i}_{h_i,\bar h_i}(u_i,z_i,\bar z_i)\>.$$ 

\smallskip

In other words, {\textit{the time-dependent correlators of a 3d Carrollian CFT compute the 4d scattering amplitudes in the Mellin basis.}} \\

We would like to emphasize that we are \textit{not} saying that every Carrollian CFT computes space-time scattering amplitude. But, if a specific Carrollian CFT does so then it does it in the modified Mellin basis \eqref{modmellin}.

\medskip

Now the reader might think that this identification is kinematical because both the objects transform in the same way under relevant symmetries. While this is correct, the dynamics enters non-trivially when we choose one of the branches of the conformal Carrollian correlation functions.

\medskip

Before we end this section we would like to emphasize few points. First of all, Celestial holography, as it stands, requires the existence of an infinite number of conformal primary fields with complex scaling dimensions. So any Carrollian CFT which can compute 4d scattering amplitudes should also have this feature. 

\medskip

The second point is regarding the symmetry group of a Carollian CFT. Over the last few years, study of tree level massless scattering amplitudes using the framework of celestial holography has revealed a much larger asymptotic symmetry group than the extended BMS$_{4}$. For example, the $SL_2$ current algebra at level zero turns out to be a symmetry algebra \cite{Banerjee:2020zlg} of tree level graviton scattering amplitudes. In fact it has been shown that the $w_{1+\infty}$ is a symmetry algebra \cite{Guevara:2021abz,Strominger:2021lvk,Himwich:2021dau,Pasterski:2022joy,Jiang:2021csc} for massless scattering amplitudes. This also holds at the loop level in some special cases. Therefore the asymptotic symmetry algebra for flat space time is expected to be far more richer than the extended BMS$_4$ algebra. The current Carrollian framework has to be extended in order to accommodate these additional symmetries.

\subsection{Correlation functions in Carrollian CFT and different branches}
Having already revealed the main punchline of our paper, we now go ahead and show the existence of two different branches of correlation functions for 3d Carroll CFTs. 

\medskip

We are interested in computing the two point vacuum correlation functions of primary fields in these 3d Carroll CFTs. We will see that just like relativistic CFTs, it is possible to completely determine (upto constant factors) the two and three-point functions using symmetry arguments. We demand that the correlation functions are invariant under the Poincare subalgebra ($\{M_{l,m}, L_n\}$ with $l,m=0,1$ and $n=0, \pm1$) of the BMS$_4$ algebra. Consider the two point function
\begin{equation}
G(u,z,\bar{z},u',z',\bar{z}')=\<0|\Phi(u,z,\bar{z})\Phi'(u',z',\bar{z}')|0\>.
\end{equation}
Here $\Phi(u,z,\bar{z})$ and $\Phi'(u',z',\bar{z}')$ are primaries with weight $(h,h')$ and $(\bar{h},\bar{h}')$ respectively. Invariance under Carroll time translations leads to 
\be{}
\left(\frac{\partial}{\partial u} + \frac{\partial}{\partial u'}\right) G(u,z,\bar z, u',z',\bar z')=0
\ee
Now considering Carroll boost invariance ($u \to u + b z + \bar{b} \z$){\footnote {3d Carroll boosts are identical to spatial translations in 4d Minkowski spacetime. For more details, see Appendices \ref{cel}, \ref{car}.}}  we get,
\begin{align}
\left(z \frac{\partial}{\partial u} + z' \frac{\partial}{\partial u'}\right) G(u,z,\bar z, u',z',\bar z')=0, \, \left(\bar z \frac{\partial}{\partial u} + \bar z' \frac{\partial}{\partial u'}\right) G(u,z,\bar z, u',z',\bar z')=0. 
\end{align}
So combining the above equations we get, 
\begin{align}
\left( z - z' \right) \frac{\partial}{\partial u} G(u - u', z-z', \bar z - \bar z') =0, \, \left( \bar z - \bar z'\right) \frac{\partial}{\partial u} G(u-u', z-z', \bar z - \bar z') =0. 
\end{align}
These equations have two independent solutions that give rise to two different classes of correlators, which has been referred to in the section above\footnote{This was noticed independently in \cite{Chen:2021xkw}.}. 

\medskip

The first class of correlators corresponds to the choice of solution
\be{}
\frac{\partial}{\partial u} G(u-u', z-z', \bar z - \bar z') = 0 
\ee
Using invariances under the subalgebra $\{L_{0,\pm1}, \L_{0,\pm1}\}$ of BMS$_4$, the above gives rise to standard 2d CFT 2-point correlation function \cite{Bagchi:2016bcd}:
\begin{equation}
G(u,z,\bar{z},u',z',\bar{z}')=\frac{\delta_{h,h'} \delta_{\h, \h'}}{(z-z')^{2h}(\bar{z}-\bar{z}')^{2\bar{h}}}.
\end{equation}
For our discussions in this paper, we will not be interested in this particular branch. The second class of solution corresponds to the choice 
\be{}
\frac{\partial}{\partial u} G(u-u', z-z', \bar z - \bar z') \propto \delta^2 \(z - z'\) 
\ee 
This gives rise to correlation functions in which we are interested \cite{Banerjee:2018gce}. Thus, 
\begin{equation} \label{f}
G(u,z,\bar{z},u',z',\bar{z}')=f(u-u')\delta^{2}(z-z').
\end{equation}
We will call this the delta-function branch{\footnote{In our notation $\delta^{2}(z) = \delta(x)\delta(y).$}}. By demanding invariance under the subalgebra $\{L_{0,\pm1}, \L_{0,\pm1}\}$ of BMS$_4$, it is straightforward to show
\be{}
(\Delta+\Delta'-2)f(u-u')+(u-u')\partial_uf(u-u')=0,  \quad (\sigma+\sigma')f(u-u')=0.
\ee
Here $\Delta=(h+\bar{h})$ is the scaling dimension and $\sigma=(h-\bar{h})$ is spin. The solution of the above equations is 
\begin{equation}
f(u-u')=C \delta_{\sigma+\sigma',0}(u-u')^{-(\Delta+\Delta'-2)},
\end{equation}
where $C$ is a constant factor. Hence
\begin{equation} \label{Sym-cor}
G(u,z,\bar{z},u',z',\bar{z}')= \frac{C \,\delta^{2}(z-z')}{(u-u')^{\Delta+\Delta'-2}} \delta_{\sigma+\sigma', 0}.
\end{equation}
Once this correlator has this form (\ref{Sym-cor}), the equation which imposes $M_{11}$ or the transformation 
$u \to u + \e z\z$ is trivially satisfied. 

\medskip

Notice that very unlike a relativistic CFT 2 point function, here one does not have to have equal scaling dimensions for the fields to get a non-zero answer. Thus this branch cannot be accessed by taking a limit from relativistic CFT correlation functions. 

\medskip

Let us now discuss how one can obtain the same two point function by modified Mellin transformation \eqref{modmellin} of scattering amplitudes \cite{Banerjee:2018gce}. Of course in the case of two point function the scattering amplitude is trivial and is given by the inner product 

\be{}
\<{p_1,\sigma_1}|{p_2,\sigma_2}\> = (2\pi)^3 2E_{p_1} \delta^3\(\vec p_1 - \vec p_2\) \delta_{\sigma_1+\sigma_2,0}
\ee
Here the notation is standard except that we label the helicity of an external particle as if it were an outgoing particle. Now using the parametrization \eqref{pmu} we can write,

\be{}
\<{p_1,\sigma_1}|{p_2,\sigma_2}\> = 4\pi^3 \frac{\delta\(\omega_1 - \omega_2\)\delta^2\(z_1 - z_2\)}{\omega_1}\delta_{\sigma_1 + \sigma_2, 0}
\ee

Now the Mellin transformed two point function is given by
\begin{align}
&\mathcal{\tilde M}\(u_1, z_1, \bar z_1, u_2, z_2, \bar z_2, h_1, \bar h_1, h_2, \bar h_2, \epsilon_1 = 1, \epsilon_2 = -1\) \nonumber\\
& = 4\pi^3 \delta_{\sigma_1 + \sigma_2,0} \int_{0}^{\infty} d\omega_1 \int_{0}^{\infty} d\omega_2 \omega_1^{\D_1 -1}\omega_2^{\D_2-1} e^{- i\omega_1 u_1}e^{i\omega_2 u_2} \frac{\delta\(\omega_1 - \omega_2\)\delta^2\(z_1 - z_2\)}{\omega_1} \nonumber\\
&= 4\pi^3 \Gamma\(\D_1 + \D_2 -2\) \frac{\delta^2\(z_1 - z_2\)}{\(i \( u_1 - u_2\)\)^{\D_1 + \D_2 -2}} \delta_{\sigma_1 + \sigma_2, 0}
\end{align}
We can see that, modulo the constant normalization, this has the same structure as the time dependent two point function of the Carrollian CFT.

\medskip

More importantly we can see that the presence of the spatial delta function $\delta^2(z_1 - z_2)$ in the Carrollian two point function has the (dual) physical interpretation that the momentum direction of a free particle in the bulk space time does not change. 

\medskip

In the same way following \cite{Banerjee:2018gce} one can show that the \textit{time dependent} three point function in the Carrollian CFT is zero. This has the physical interpretation that in Minkowski signature the scattering amplitude of three massless particles vanish due to energy momentum conservation. 

\medskip

Therefore, \textit{the peculiarities of the time dependent correlation functions of a Carrollian CFT are precisely what we need to connect to the spacetime scattering amplitudes of massless particles}. 

\medskip

This is the main message of our paper. 

\medskip

\section{Massless Carrollian scalar field}
In this section, our primary goal is to provide a concrete example of a 3d quantum field theory which is invariant under the BMS$_4$ algebra and give us the correlation functions in the delta-function branch. We will now focus on a particular simple example, that of a massless Carroll scalar field.

\subsection*{The Action}

The minimally coupled massless scalar field on Carrollian  backgrounds is described by 
\begin{equation}\label{covact}
	\mathcal{S}=\int du d^{2}x^i \, \tau^\mu \tau^\nu  \partial_{\mu}\Phi\partial_{\nu}\Phi
\end{equation}
where $\tau^\mu$ are vector fields defined in Eq. \refb{gtau}. We shall work with flat Carroll background by fixing $\tau^\mu =(1,0)$ and $g_{ij}=\delta_{ij}$. On flat Carroll background the action takes the simple form
\begin{equation} \label{lag}
	\mathcal{S}=\int dud^2x^i  \ \frac{1}{2} (\p_u{\Phi})^2
\end{equation}
The two dimensional version of this action \refb{covact} has been extensively used to study the tensionless limit of string theory where the BMS$_3$ algebra replaces the two copies of the Virasoro algebra as worldsheet symmetries (see e.g. \cite{Bagchi:2015nca}). We will see that here this simple 3d action carries the seeds of a potential dual formulation of 4d gravity in asymptotically flat spacetimes. The above action  \refb{covact} also arises as the leading piece in the Carroll expansion (speed of light $c\to0$) of the action of the free relativistic massless scalar field theory in (2+1) dimensions.

\subsection*{Two-point function of the Carroll scalar}
Now we turn our attention to the construction of the two point correlation function of the Carroll scalar. We shall do this in two ways, first by a Green's function method and then by canonical analysis. We will see that we will end up on the delta-function branch elucidated by our general analysis in the previous sections. 

\subsubsection*{Green's function} 
We first compute the two point correlation function of the massless Carroll scalar by computing the Green's functions. The Green's function equation for this theory would be
\begin{equation}
	\partial^2_u G(u-u, z^i-z'^i)=\delta^{3}(u-u',z^i-z'^i).
\end{equation} 
This equation can be solved in usual way by going to Fourier space where this takes the form  
\begin{equation}
	\tilde{G}(k_u,k_i)=-\frac{1}{k_u^2}.
\end{equation}
Transforming back into position space yields 
\begin{align}
	G(u-u',z^i-z'^i) & =-\int \frac{dk_u}{k_u^2+\mu^2}e^{ik_u(u-u')}\int d^2 z\  e^{ik_i(z^i-z'^i)} \nonumber \\
	&=\frac{i}{2}\left[\frac{1}{\mu}-(u-u')\right]\delta^{(2)}(z^i-z'^i).
\end{align}
As the equation of motion does not have any spatial derivatives this integral diverges. We regulate this integral by throwing away the troublesome infinite piece: 
\begin{equation}
	G(u-u',z^i-z'^i)=-\frac{i}{2}(u-u')\delta^{2}(z-z',\bar{z}-\bar{z}'))
\end{equation}
For scalar fields,  the spin $\sigma=0$ and conformal weights of $\Phi(u,z,\bar{z})$ are (e.g. from the action \refb{lag}) 
\be{}
h=\frac{1}{4}, \quad \bar{h}=\frac{1}{4}. 
\ee
Hence the answer obtained by the Green's function method is in perfect agreement with the 2-point function derived from symmetry arguments in the earlier section.

\subsubsection*{Canonical approach}
We will now rederive the scalar two point correlation function by taking recourse to cannonical quantisation. To proceed we will put the free scalar theory on the round sphere and then take radius to infinity limit to recover our answer in the plane coordinates. The scalar field action on a manifold with topology $\mathbb{R} \times \mathbb{S}^2$ would be 
 \begin{equation}
 \mathcal{S}=\int du d^2 z \sqrt{q} \left[\frac{1}{2}(\p_u\Phi)^2-k^2\Phi^2\right]
 \end{equation}
Here $k$ is related to the radius of the sphere $R$ by $k=\frac{1}{2R}$. The metric on the sphere is denoted by $q_{ij}$ and is given by \refb{q}. The Euler-Lagrange equations of motion for this action is 
 \begin{equation}
 \ddot{\Phi}+k^2\Phi^2=0
 \end{equation}
Generic real solutions are given by: 
\be{}
\Phi(u,z,\bar{z})=\frac{1}{\sqrt{k}} \left( C^\dagger(z,\bar{z})e^{iku}+C (z, \bar{z})e^{-iku}\right).
\ee
The canonical commutation relation between the $C$ fields are and the Hamiltonian are respectively:
\begin{align}
& [C (z, \bar{z}), C^{\dagger}(z', \bar{z}')]=\frac{1}{2} \delta^2(z-z'), \\
&H=k \int d^2 z \sqrt{q}  \left(  2  C^{\dagger}(z, \bar{z}) C(z, \bar{z}) + \frac{1}{2} \delta^2(0) \right).
\end{align}
The Hamiltonian has the unphysical zero point energy in the form of the $\delta^2 (0)$. The physical part of the Hamiltonian then implies that the time translation symmetric ground state should be annihilated by $C$:  
\be{}
C(z, \bar{z}) | 0 \rangle = 0,\, \mbox{ for } \, (z,  \bar{z}) \in \mathbb{S}^2.
\ee

It is therefore straightforward to calculate the 2-point function (keeping in mind that there are no zero modes):
\begin{align}
G(u,u',z^i,z'^i)=\<0|T\Phi(u,z,\bar{z})\Phi(u',z',\bar{z}')|0\>.
\end{align}
Now, taking $u>u'$, we obtain
\begin{align}
G(u,u',z^i,z'^i)= -\frac{1}{2k}[\cos k(u-u')+i \sin k(u-u')]\delta^{2}(z-z',\bar{z}-\bar{z}')
\end{align}
So in the limit $R \to \infty$ , or $k \to 0$ , the two point function becomes 
\begin{equation} \label{phys2pt}
G(u,u',z^i,z'^i)=-\left[\frac{1}{2k}+\frac{i}{2} (u-u')\right]\delta^{2}(z-z',\bar{z}-\bar{z}')
\end{equation}
We read of the scaling dimension of $\Phi$ as $\Delta = 1/2$, since this is spin-less, the physical part of the 2 point function \eqref{phys2pt} matches exactly with the one computed using symmetry arguments.

\section{Discussions}
In this paper, we have provided evidence that the correlation functions of 3d Carrollian CFTs encode scattering amplitudes for 4d asympotically flat spacetimes, specifically in the Mellin basis. The Carroll correlators have two distinct branches and one of these, the one with explicit Carroll time dependence which we called the delta-function branch, was the one relevant for the connection to flat space scattering. 

\medskip

There are a number of intriguing questions that arise from our considerations in this paper. Originally, the version of flat space holography that was envisioned with the connections with Carroll CFTs was one which emerged as a systematic limit from AdS/CFT. It is clear that the correlation functions that we focused on in this work cannot emerge as a Carroll limit from standard relativistic 3d CFT correlators in position space, since e.g. the CFT 2-point function would vanish for unequal weight primaries and in the time-dependent Carroll branch, this does not happen. Hence it would seem that the formulation of Carroll holography we would require for connection to scattering amplitudes would be disconnected from AdS/CFT. While this makes sense because flat space and AdS are fundamentally different, how this fits in with e.g. the programme of attempting to find flat space correlations from AdS/CFT (see e.g. \cite{Giddings:1999jq,Gary:2009mi,Raju:2012zr}), remains to be seen. 

\medskip

Our construction and specifically the emergence of two different branches of correlation functions is also reminiscent of recent advances in the tensionless regime of string theory where three distinct quantum theories appear from a single classical theory \cite{Bagchi:2020fpr,Bagchi:2021rfw}. Recent findings of different correlation functions in these theories \cite{Sud,Hao:2021urq} is an indication that perhaps there is an interesting non-trivial quantum vacuum structure underlying the Carrollian theories we have discussed in our work. 

\medskip

It would be of interest to figure out what, if any, is relation of the 2d CFT branch of the Carroll correlation functions to 4d flatspace physics. We would also like to understand if it is possible to construct explicit examples of 3d Carroll CFTs exhibiting 2d CFT correlation functions.  

\medskip

As mentioned before, in order to reproduce flatspace scattering, as in Celestial CFTs, Carroll CFTs need an infinite number of primary fields with complex scaling dimension. Moreover, the celestial CFT enjoys a much larger symmetry \cite{Banerjee:2020zlg,Banerjee:2020vnt,Guevara:2021abz,Strominger:2021lvk,Himwich:2021dau,Pasterski:2022joy,Jiang:2021csc,Campiglia:2014yka,Banerjee:2021dlm,Donnay:2020guq,Freidel:2021ytz,Freidel:2021dfs} than the extended BMS$_4$ algebra and these additional symmetries play a central role in holographic computation of scattering amplitudes. We would ideally like to construct an explicit example of such a theory to provide a concrete proposal between a gravitational theory in asymptotically flat spacetimes and a Carroll CFT with additional symmetries living on its null boundary. But it is obvious that this is presently a distant goal. 

\bigskip

\subsection*{Acknowledgements}
We thank Kinjal Banerjee and Amartya Saha for discussions and Daniel Grumiller for helpful correspondence. 

\smallskip

AB and SB are partially supported by Swarnajayanti Fellowships of the Department of Science and Technology, India. AB is also partially supported by Science and Engineering Research Board (SERB) grants SB/SJF/2019-20/08, CRG/2020/002035. The work of SB is also partially supported by SERB grant MTR/2019/000937 (Soft-Theorems, S-matrix and Flat-Space Holography). RB is also supported by the following grants from the SERB: CRG/2020/002035, SRG/2020/001037. SD is supported by grant number 09/092(0971)/2017-EMR-I from Council of Scientific and Industrial Research (CSIR).

\bigskip \bigskip
\section*{APPENDICES}

\bigskip 
\appendix

\section{Brief review of Celestial or Mellin amplitudes for massless particles}\label{cel}
In this appendix, we provide a lightning review of Celestial amplitudes for massless particles for ready reference. The Celestial or Mellin amplitude for massless particles in four dimensions is defined as the Mellin transformation of the $S$-matrix element, given by \cite{Pasterski:2016qvg,Pasterski:2017kqt}
\be{mellin}
\mathcal M_n\big(\{z_i, \bar z_i, h_i, \bar h_i\}\big) = \prod_{i=1}^{n} \int_{0}^{\infty} d\omega_i \ \omega_i^{\D_i -1} S_n\big(\{\omega_i,z_i,\bar z_i, \sigma_i\}\big)
\ee 
where $\sigma_i$ denotes the helicity of the $i$-th particle and the on-shell momenta are parametrized as,
\be{para}
p_i = \omega_i (1+z_i\bar z_i, z_i + \bar z_i , -i(z_i - \bar z_i), 1- z_i \bar z_i), \quad p_i^2 = 0
\ee
The scaling dimensions $(h_i,\bar h_i)$ are defined as,
\be{}
h_i = \frac{\D_i + \sigma_i}{2}, \quad \bar h_i = \frac{\D_i - \sigma_i}{2}
\ee
The Lorentz group $SL(2,\mathbb C)$ acts on the celestial sphere as the group of global conformal transformations and the Mellin amplitude $\mathcal M_n$ transforms as,
\be{}
\mathcal M_n\big(\{z_i, \bar z_i, h_i, \bar h_i\}\big) = \prod_{i=1}^{n} \frac{1}{(cz_i + d)^{2h_i}} \frac{1}{(\bar c \bar z_i + \bar d)^{2\bar h_i}} \mathcal M_n\bigg(\frac{az_i+b}{cz_i+d} \ ,\frac{\bar a \bar z_i + \bar b}{\bar c \bar z_i + \bar d} \ , h_i,\bar h_i\bigg)
\ee
This is the familiar transformation law for the correlation function of primary operators of weight $(h_i,\bar h_i)$ in a $2$-D CFT under the global conformal group $SL(2,\mathbb C)$.
\vskip 4pt
In Einstein gravity, the Mellin amplitude as defined in \eqref{mellin} usually diverges. This divergence can be regulated by defining a modified Mellin amplitude as \cite{Banerjee:2018gce,Banerjee:2018fgd}, 
\be{mellinmod}
\mathcal M_n\big(\{u_i,z_i, \bar z_i, h_i, \bar h_i\}\big) = \prod_{i=1}^{n} \int_{0}^{\infty} d\omega_i \ \omega_i^{\D_i -1} e^{-i\sum_{i=1}^n \epsilon_i \omega_i u_i} S_n\big(\{\omega_i,z_i,\bar z_i, \sigma_i\}\big)
\ee 
where $u$ can be thought of as a time coordinate and $\epsilon_i = \pm 1$ for an outgoing (incoming) particle. Under (Lorentz) conformal tranansformation the modified Mellin amplitude $\mathcal M_n$ transforms as,
\be{M}
\mathcal M_n\big(\{u_i,z_i, \bar z_i, h_i, \bar h_i\}\big) = \prod_{i=1}^{n} \frac{1}{(cz_i + d)^{2h_i}} \frac{1}{(\bar c \bar z_i + \bar d)^{2\bar h_i}} \mathcal M_n\bigg(\frac{u_i}{|cz_i + d|^2} \ , \frac{az_i+b}{cz_i+d} \ ,\frac{\bar a \bar z_i + \bar b}{\bar c \bar z_i + \bar d} \ , h_i,\bar h_i\bigg)
\ee
Under global space-time translation, $u \rightarrow u + A + Bz + \bar B\bar z + C z\bar z$, the modified amplitude is invariant, i.e, 
\be{}
\mathcal M_n\big(\{u_i + A + Bz_i + \bar B\bar z_i + C z_i\bar z_i ,z_i, \bar z_i, h_i, \bar h_i\}\big) = \mathcal M_n\big(\{u_i,z_i, \bar z_i, h_i, \bar h_i\}\big)
\ee
Now in order to make manifest the conformal nature of the dual theory living on the celestial sphere it is useful to write the (modified) Mellin amplitude as a correlation function of conformal primary operators. So let us define a generic conformal primary operator as, 
\be{confprim}
\phi^{\epsilon}_{h,\bar h}(z,\bar z) = \int_{0}^{\infty} d\omega \  \omega^{\D-1} a(\epsilon\omega, z, \bar z, \sigma)
\ee
where $\epsilon=\pm 1$ for an annihilation (creation) operator of a massless particle of helicity $\sigma$. Under (Lorentz) conformal transformation the conformal primary transforms like a primary operator of scaling dimension $(h,\bar h)$
\be{}
\phi'^{\epsilon}_{h,\bar h}(z,\bar z) = \frac{1}{(cz + d)^{2h}} \frac{1}{(\bar c \bar z + \bar d)^{2\bar h}} \mathcal \phi^{\epsilon}_{h,\bar h}\bigg(\frac{az+b}{cz+d} \ ,\frac{\bar a \bar z + \bar b}{\bar c \bar z + \bar d}\bigg)
\ee
Similarly in the presence of the time coordinate $u$ we have,
\be{confprimu}
\phi^{\epsilon}_{h,\bar h}(u,z,\bar z) = \int_{0}^{\infty} d\omega \ \omega^{\D-1} e^{-i \epsilon \omega u} a(\epsilon\omega, z, \bar z, \sigma)
\ee
Under (Lorentz) conformal transformations 
\be{}
\phi'^{\epsilon}_{h,\bar h}(u,z,\bar z) = \frac{1}{(cz + d)^{2h}} \frac{1}{(\bar c \bar z + \bar d)^{2\bar h}} \mathcal \phi^{\epsilon}_{h,\bar h}\bigg(\frac{u}{|cz+d|^2},\frac{az+b}{cz+d} \ ,\frac{\bar a \bar z + \bar b}{\bar c \bar z + \bar d}\bigg)
\ee
In terms of \eqref{confprim},  the Mellin amplitude can be written as the correlation function of conformal primary operators
\be{}
\mathcal M_n = \bigg\langle{\prod_{i=1}^n \phi^{\epsilon_i}_{h_i,\bar h_i}(z_i,\bar z_i)}\bigg\rangle
\ee
Similarly using \eqref{confprimu}, the modified Mellin amplitude can be written as,
\be{}
\mathcal M_n = \bigg\langle{\prod_{i=1}^n \phi^{\epsilon_i}_{h_i,\bar h_i}(u_i,z_i,\bar z_i)}\bigg\rangle
\ee

\bigskip 
\bigskip

\section{Conformal Carroll symmetry}\label{car}
We will briefly summarize the main points relating to Conformal Carroll symmetry in this appendix. 

\subsection*{Geometric structures}
Throughout the paper, we have been interested in defining quantum field theories that live on the null boundary $\mathscr{I}^\pm$ of asymptotically flat spacetimes. The null boundary is topologically $\mathbb{R}_u \times \mathbb{S}^2$ with $\mathbb{R}_u$ being null. The induced metric of $\mathscr{I}^\pm$  is degenerate and Carroll structures replace usual Riemann structures on it. Carrollian manifolds are endowed with a degenerate, twice symmetric metric $g_{\mu\nu}$ and its kernel vector field $\tau^\mu$. The isometry algebra of this structure 
\be{gtau}
\mathcal{L}_\xi \tau^\mu = 0, \quad \mathcal{L}_\xi g_{\mu\nu} =0, \quad \tau^\mu g_{\mu\nu} =0
\ee
generates the Carroll algebra, which for flat Carroll manifolds reduces to the group that is obtained from the Poincare group by sending the speed of light to zero \cite{Bagchi:2016bcd}. We shall see the algebra and the contraction below. 

\medskip

We have dealt with CFTs on these Carroll backgrounds. These theories are naturally expected to be invariant under the conformal isometries of these Carrollian structures. Conformal Killing equations on these background are 
\begin{eqnarray}
	\mathcal{L}_{\xi}g_{\mu\nu}=\lambda g_{\mu\nu}, \quad \mathcal{L}_{\xi}\tau^\mu=-\frac{\lambda}{N}\tau^\mu
\end{eqnarray}
The so called dynamical exponent $N$ encapsulates the relative scaling of space and time directions. For $N=2$, the space and time dilates homogeneously. This is the case that we will be interested, as we wish to connect the conformal Carroll structures to the asymptotic structure of flat spacetimes, where of course space and time scale in the same way. In (2+1) dimensions, the set of vector fields that solves the above equation for $N=2$ is 
\begin{equation}
\xi=\left[\alpha(x^i)+\frac{u}{2{\sqrt{q}}}\partial_{i}({\sqrt{q}} f^i(x^j))\right]\partial_u+f^{i}(x^j)\partial_i \label{kill}
\end{equation}
Here $q$ is the determinant of the metric $q_{ij}$ on $\mathbb{S}^2$ and $\alpha(x^i)$ is an arbitrary function of $x^i$, but $f^{i}(x^i)$ need to satisfy the following conformal Killing equation on $\mathbb{S}^2$:
\begin{eqnarray}
\mathcal{L}_f q_{ij} = D_k f^k \, q_{ij}
\end{eqnarray}
$D$ is the connection compatible with $q_{ij}$. Choosing the stereographic coordinates $(z, \bar{z})$ on 2 sphere, such that 
\be{q}
ds^2 = \dfrac{2 dz \, d \bar{z}}{(1+z \bar{z})^2},
\ee 
the above equation for the components of $f$ is solved by holomorphic and anti-holomorphic functions, i.e $f^z \equiv f^z(z)$ and $f^{\bar{z}} \equiv f^{\bar{z}}(\bar{z})$. The algebra of these vector fields \eqref{kill} is clearly infinite dimensional and interestingly closes to form the BMS$_4$ algebra \refb{bms4} \cite{Duval:2014lpa}. 

The connection between Carrollian conformal symmetries and BMS symmetries for arbitrary dimensions was clarified in \cite{Duval:2014lpa}, following closely related observations in \cite{Bagchi:2010zz}.

\subsection*{Carroll contractions of Conformal algebra}

The Carrollian limit of a relativistic CFT is reached by performing  an In\"{o}n\"{u}--Wigner
contraction on the relativistic conformal generators. The corresponding contraction of the spacetime coordinates for a 3d CFT is described as
\be{stscale}
x_i \to x_i, \qquad t \to \e\, t, \qquad \e \to 0\, \quad \text{with} \quad i = 1,2.
\ee
This is equivalent to taking the speed of light, $c\to 0$. Conformal Carroll generators are obtained from their relativistic counterparts in this limit:
\begin{subequations}
\label{genearl}
\begin{align}
H &= \p_t &  B_i&=-x_i \p_t &  K_i &= -2 x_j (t\p_t+x_i\p_i)+x_j x_j \p_i & K &=x_i x_i \p_t \\
 D&=-(t\p_t+x_i \p_i) & P_i&=\p_i & J_{ij}&=-(x_i\p_j-x_j\p_i)\,. &&
\end{align} 
\end{subequations}
These generate the 3d finite Conformal Carrollian algebra which is $iso(3,1)$ and hence isomorphic to the Poincare group in $d=4$ \cite{Bagchi:2016bcd}. The contracted algebra is given by
\begin{align}
[J_{ij}, B_k ]&=\delta_{k[i}B_{j]} & [J_{ij}, P_k ]&=\delta_{k[i}P_{j]} & [J_{ij}, K_k ]&=\delta_{k[i}K_{j]} & [B_i,P_j]&=\delta_{ij}H\nonumber\\
[B_i,K_j]&=\delta_{ij}K & [D,K]&=-K & [K,P_i]&=2B_i & [K_i,P_j]&=-2D\delta_{ij}-2J_{ij} \nonumber \\ [H,K_i]&=2B_i & [D,H]&=H & [D,P_i]&=P_i & [D,K_i]&=-K_i\,.
\end{align}
The sub-algebra consisting of the generators $\{J_{ij}, B_i, P_i, H\}$ forms the Carrollian algebra, the $c\to0$ limit of the 3d Poincar{\'e} algebra. The generators $\{J_{ij},P_i,D,K_i\}$ form the conformal algebra of celestial sphere or equivalently the 4d Lorentz algebra. 

\subsection*{3d Finite Conformal Carroll = 4d Poincare}
One of the most crucial observations that our work is based on is the fact the 3d finite Conformal Carroll algebra is isomorphic to the 4d Poincare algebra. Here we summarise the relation below:
\begin{subequations}
\begin{align}
&M_{00} = H, \quad M_{01} =  B_x +i B_y, \quad  M_{10} =  B_x -i B_y, \quad M_{11} = K_0 \\
&L_0=\frac{1}{2}(D+ iJ), \quad L_{-1}=\frac{1}{2}(P_x+ iP_y)  \quad L_{1}=\frac{1}{2}(K_x+ iK_y) \\
&\L_0=\frac{1}{2}(D- iJ) \quad \L_{-1}=\frac{1}{2}(P_x- iP_y) \quad \L_{1}=\frac{1}{2}(K_x- iK_y)
\end{align}
\end{subequations}
Here the LHS of the equations are the 4d Poincare generators and the RHS are the finite Conformal Carroll generators. We see that the 4d spacetime translations $M_{r,s}$ arrange themselves into the Hamiltonian, Carroll boosts, and the temporal part of the Carroll SCT. The Lorentz generators $L_n, \L_n$ become the global conformal generators on the sphere, as expected.


\newpage


\begin{thebibliography}{999}

\bibitem{Bondi:1962px} 
  H.~Bondi, M.~G.~J.~van der Burg and A.~W.~K.~Metzner,
  ``Gravitational waves in general relativity. 7. Waves from axisymmetric isolated systems,''
  Proc.\ Roy.\ Soc.\ Lond.\ A {\bf 269}, 21 (1962).

  R.~Sachs,
  ``Asymptotic symmetries in gravitational theory,''
  Phys.\ Rev.\  {\bf 128}, 2851 (1962).


\bibitem{Strominger:2013jfa}
A.~Strominger,
``On BMS Invariance of Gravitational Scattering,''
JHEP \textbf{07}, 152 (2014)
doi:10.1007/JHEP07(2014)152
[arXiv:1312.2229 [hep-th]].


\bibitem{He:2014laa}
T.~He, V.~Lysov, P.~Mitra and A.~Strominger,
``BMS supertranslations and Weinberg\textquoteright{}s soft graviton theorem,''
JHEP \textbf{05}, 151 (2015)
doi:10.1007/JHEP05(2015)151
[arXiv:1401.7026 [hep-th]].

\bibitem{Strominger:2014pwa}
A.~Strominger and A.~Zhiboedov,
``Gravitational Memory, BMS Supertranslations and Soft Theorems,''
JHEP \textbf{01}, 086 (2016)
doi:10.1007/JHEP01(2016)086
[arXiv:1411.5745 [hep-th]].

\bibitem{Barnich:2009se}
G.~Barnich and C.~Troessaert,
``Symmetries of asymptotically flat 4 dimensional spacetimes at null infinity revisited,''
Phys. Rev. Lett. \textbf{105}, 111103 (2010)
doi:10.1103/PhysRevLett.105.111103
[arXiv:0909.2617 [gr-qc]].

\bibitem{Barnich:2010eb}
G.~Barnich and C.~Troessaert,
``Aspects of the BMS/CFT correspondence,''
JHEP \textbf{05}, 062 (2010)
doi:10.1007/JHEP05(2010)062
[arXiv:1001.1541 [hep-th]].


\bibitem{Cachazo:2014fwa}
F.~Cachazo and A.~Strominger,
``Evidence for a New Soft Graviton Theorem,''
[arXiv:1404.4091 [hep-th]].


 \bibitem{Kapec:2014opa} 
  D.~Kapec, V.~Lysov, S.~Pasterski and A.~Strominger,
  ``Semiclassical Virasoro symmetry of the quantum gravity $ \mathcal{S}$-matrix,''
  JHEP {\bf 1408}, 058 (2014)
  doi:10.1007/JHEP08(2014)058
  [arXiv:1406.3312 [hep-th]].
  
  \bibitem{Kapec:2016jld}
  D.~Kapec, P.~Mitra, A.~M.~Raclariu and A.~Strominger,
  ``2D Stress Tensor for 4D Gravity,''
  Phys.\ Rev.\ Lett.\  {\bf 119}, no. 12, 121601 (2017)
  doi:10.1103/PhysRevLett.119.121601
  [arXiv:1609.00282 [hep-th]].
  
  \bibitem{He:2017fsb} 
  T.~He, D.~Kapec, A.~M.~Raclariu and A.~Strominger,
  ``Loop-Corrected Virasoro Symmetry of 4D Quantum Gravity,''
  JHEP {\bf 1708}, 050 (2017)
  doi:10.1007/JHEP08(2017)050
  [arXiv:1701.00496 [hep-th]].
 
  \bibitem{Ball:2019atb} 
  A.~Ball, E.~Himwich, S.~A.~Narayanan, S.~Pasterski and A.~Strominger,
  ``Uplifting AdS3/CFT2 to Flat Space Holography,''
 arXiv:1905.09809 [hep-th].
 
 \bibitem{Kapec:2017gsg} 
D.~Kapec and P.~Mitra,
``A $d$-Dimensional Stress Tensor for Mink$_{d+2}$ Gravity,''
arXiv:1711.04371 [hep-th]. 

\bibitem{Fotopoulos:2019vac}
A.~Fotopoulos, S.~Stieberger, T.~R.~Taylor and B.~Zhu,
``Extended BMS Algebra of Celestial CFT,''
JHEP \textbf{03}, 130 (2020)
doi:10.1007/JHEP03(2020)130
[arXiv:1912.10973 [hep-th]]

\bibitem{Stieberger:2018onx}
S.~Stieberger and T.~R.~Taylor,
``Symmetries of Celestial Amplitudes,''
Phys. Lett. B \textbf{793}, 141-143 (2019)
doi:10.1016/j.physletb.2019.03.063
[arXiv:1812.01080 [hep-th]]


  
   \bibitem{deBoer:2003vf} 
  J.~de Boer and S.~N.~Solodukhin,
  ``A Holographic reduction of Minkowski space-time,''
  Nucl.\ Phys.\ B {\bf 665}, 545 (2003)
  doi:10.1016/S0550-3213(03)00494-2
  [hep-th/0303006]. 
  
   \bibitem{Pasterski:2016qvg} 
  S.~Pasterski, S.~H.~Shao and A.~Strominger,
  ``Flat Space Amplitudes and Conformal Symmetry of the Celestial Sphere,''
  Phys.\ Rev.\ D {\bf 96}, no. 6, 065026 (2017)
  doi:10.1103/PhysRevD.96.065026
  [arXiv:1701.00049 [hep-th]].

  \bibitem{Pasterski:2017kqt} 
  S.~Pasterski and S.~H.~Shao,
  ``Conformal basis for flat space amplitudes,''
  Phys.\ Rev.\ D {\bf 96}, no. 6, 065022 (2017)
  doi:10.1103/PhysRevD.96.065022
  [arXiv:1705.01027 [hep-th]].  
  
  \bibitem{Banerjee:2018gce} 
  S.~Banerjee,
  ``Null Infinity and Unitary Representation of The Poincare Group,''
  JHEP {\bf 1901}, 205 (2019)
  doi:10.1007/JHEP01(2019)205
  [arXiv:1801.10171 [hep-th]].
  
 \bibitem{Banerjee:2018fgd} 
  S.~Banerjee,
  ``Symmetries of free massless particles and soft theorems,''
  Gen.\ Rel.\ Grav.\  {\bf 51}, no. 9, 128 (2019)
  doi:10.1007/s10714-019-2609-z
  [arXiv:1804.06646 [hep-th]].
  
  \bibitem{Banerjee:2020zlg}
S.~Banerjee, S.~Ghosh and P.~Paul,
``MHV Graviton Scattering Amplitudes and Current Algebra on the Celestial Sphere,''
[arXiv:2008.04330 [hep-th]].

\bibitem{Banerjee:2020vnt}
S.~Banerjee and S.~Ghosh,
``MHV gluon scattering amplitudes from celestial current algebras,''
JHEP \textbf{10}, 111 (2021)
doi:10.1007/JHEP10(2021)111
[arXiv:2011.00017 [hep-th]].

 \bibitem{Guevara:2021abz}
A.~Guevara, E.~Himwich, M.~Pate and A.~Strominger,
``Holographic Symmetry Algebras for Gauge Theory and Gravity,''
[arXiv:2103.03961 [hep-th]].

\bibitem{Strominger:2021lvk}
A.~Strominger,
``w(1+infinity) and the Celestial Sphere,''
[arXiv:2105.14346 [hep-th]]

\bibitem{Himwich:2021dau}
E.~Himwich, M.~Pate and K.~Singh,
``Celestial operator product expansions and w$_{1+\infty}$ symmetry for all spins,''
JHEP \textbf{01}, 080 (2022)
doi:10.1007/JHEP01(2022)080
[arXiv:2108.07763 [hep-th]]

\bibitem{Pasterski:2022joy}
S.~Pasterski and H.~Verlinde,
``Mapping SYK to the Sky,''
[arXiv:2201.05054 [hep-th]]

\bibitem{Freidel:2021ytz}
L.~Freidel, D.~Pranzetti and A.~M.~Raclariu,
``Higher spin dynamics in gravity and $w_{1 + \infty}$ celestial symmetries,''
[arXiv:2112.15573 [hep-th]].

\bibitem{Freidel:2021dfs}
L.~Freidel, D.~Pranzetti and A.~M.~Raclariu,
``Sub-subleading Soft Graviton Theorem from Asymptotic Einstein's Equations,''
[arXiv:2111.15607 [hep-th]].

\bibitem{Jiang:2021csc}
H.~Jiang,
``Celestial OPEs and w$_{1+\infty}$ algebra from worldsheet in string theory,''
JHEP \textbf{01}, 101 (2022)
doi:10.1007/JHEP01(2022)101
[arXiv:2110.04255 [hep-th]]

\bibitem{Donnay:2020guq}
L.~Donnay, S.~Pasterski and A.~Puhm,
``Asymptotic Symmetries and Celestial CFT,''
[arXiv:2005.08990 [hep-th]].

\bibitem{Campiglia:2014yka}
M.~Campiglia and A.~Laddha,
``Asymptotic symmetries and subleading soft graviton theorem,''
Phys. Rev. D \textbf{90}, no.12, 124028 (2014)
doi:10.1103/PhysRevD.90.124028
[arXiv:1408.2228 [hep-th]].

\bibitem{Compere:2018ylh}
G.~Comp\`ere, A.~Fiorucci and R.~Ruzziconi,
``Superboost transitions, refraction memory and super-Lorentz charge algebra,''
JHEP \textbf{11}, 200 (2018)
[erratum: JHEP \textbf{04}, 172 (2020)]
doi:10.1007/JHEP11(2018)200
[arXiv:1810.00377 [hep-th]].

\bibitem{Banerjee:2021dlm}
S.~Banerjee, S.~Ghosh and P.~Paul,
``(Chiral) Virasoro invariance of the tree-level MHV graviton scattering amplitudes,''
[arXiv:2108.04262 [hep-th]].

\bibitem{Banerjee:2020kaa}
S.~Banerjee, S.~Ghosh and R.~Gonzo,
``BMS symmetry of celestial OPE,''
JHEP \textbf{04}, 130 (2020)
doi:10.1007/JHEP04(2020)130
[arXiv:2002.00975 [hep-th]].

\bibitem{Banerjee:2019prz}
S.~Banerjee, S.~Ghosh, P.~Pandey and A.~P.~Saha,
``Modified celestial amplitude in Einstein gravity,''
JHEP \textbf{03}, 125 (2020)
doi:10.1007/JHEP03(2020)125
[arXiv:1909.03075 [hep-th]]


\bibitem{Strominger:2017zoo}
A.~Strominger,
``Lectures on the Infrared Structure of Gravity and Gauge Theory,''
[arXiv:1703.05448 [hep-th]].

\bibitem{Pasterski:2021rjz}
S.~Pasterski,
``Lectures on celestial amplitudes,''
Eur. Phys. J. C \textbf{81}, no.12, 1062 (2021)
doi:10.1140/epjc/s10052-021-09846-7
[arXiv:2108.04801 [hep-th]].

\bibitem{Raclariu:2021zjz}
A.~M.~Raclariu,
``Lectures on Celestial Holography,''
[arXiv:2107.02075 [hep-th]].


\bibitem{Barnich:2006av} 
  G.~Barnich and G.~Compere,
  ``Classical central extension for asymptotic symmetries at null infinity in three spacetime dimensions,''
  Class.\ Quant.\ Grav.\  {\bf 24}, F15 (2007)
  [gr-qc/0610130].

\bibitem{Bagchi:2010zz}
A.~Bagchi,
``Correspondence between Asymptotically Flat Spacetimes and Nonrelativistic Conformal Field Theories,''
Phys. Rev. Lett. \textbf{105}, 171601 (2010)
doi:10.1103/PhysRevLett.105.171601
[arXiv:1006.3354 [hep-th]].


\bibitem{Bagchi:2012cy} 
  A.~Bagchi and R.~Fareghbal,
  ``BMS/GCA Redux: Towards Flatspace Holography from Non-Relativistic Symmetries,''
  JHEP {\bf 1210}, 092 (2012)
  [arXiv:1203.5795 [hep-th]].


\bibitem{Bagchi:2012xr} 
  A.~Bagchi, S.~Detournay, R.~Fareghbal and J.~Simon,
  ``Holography of 3d Flat Cosmological Horizons,''
  arXiv:1208.4372 [hep-th].

\bibitem{Barnich:2012xq} 
  G.~Barnich,
  ``Entropy of three-dimensional asymptotically flat cosmological solutions,''
  JHEP {\bf 1210}, 095 (2012)
  [arXiv:1208.4371 [hep-th]].

\bibitem{Bagchi:2013qva}
A.~Bagchi and R.~Basu,
``3D Flat Holography: Entropy and Logarithmic Corrections,''
JHEP \textbf{03}, 020 (2014)
doi:10.1007/JHEP03(2014)020
[arXiv:1312.5748 [hep-th]].

\bibitem{Bagchi:2015wna}
A.~Bagchi, D.~Grumiller and W.~Merbis,
``Stress tensor correlators in three-dimensional gravity,''
Phys. Rev. D \textbf{93}, no.6, 061502 (2016)
doi:10.1103/PhysRevD.93.061502
[arXiv:1507.05620 [hep-th]].

  \bibitem{Bagchi:2014iea} 
  A.~Bagchi, R.~Basu, D.~Grumiller and M.~Riegler,
  ``Entanglement entropy in Galilean conformal field theories and flat holography,''
  Phys.\ Rev.\ Lett.\  {\bf 114}, no. 11, 111602 (2015)
  [arXiv:1410.4089 [hep-th]].
  
\bibitem{Jiang:2017ecm}
H.~Jiang, W.~Song and Q.~Wen,
``Entanglement Entropy in Flat Holography,''
JHEP \textbf{07}, 142 (2017)
doi:10.1007/JHEP07(2017)142
[arXiv:1706.07552 [hep-th]].

\bibitem{Hijano:2017eii}
E.~Hijano and C.~Rabideau,
``Holographic entanglement and Poincar\'e blocks in three-dimensional flat space,''
JHEP \textbf{05}, 068 (2018)
doi:10.1007/JHEP05(2018)068
[arXiv:1712.07131 [hep-th]].

\bibitem{Barnich:2012aw} 
  G.~Barnich, A.~Gomberoff and H.~A.~Gonzalez,
  ``The Flat limit of three dimensional asymptotically anti-de Sitter spacetimes,''
  Phys.\ Rev.\ D {\bf 86}, 024020 (2012)
  [arXiv:1204.3288 [gr-qc]].

\bibitem{Bagchi:2012yk} 
  A.~Bagchi, S.~Detournay and D.~Grumiller,
  ``Flat-Space Chiral Gravity,''
  Phys.\ Rev.\ Lett.\  {\bf 109}, 151301 (2012)
  [arXiv:1208.1658 [hep-th]].

\bibitem{Duval:2014lpa} 
  C.~Duval, G.~W.~Gibbons and P.~A.~Horvathy,
  ``Conformal Carroll groups,''
  J.\ Phys.\ A {\bf 47}, 335204 (2014)
  [arXiv:1403.4213 [hep-th]].



\bibitem{Hartong:2015usd}
J.~Hartong,
``Holographic Reconstruction of 3D Flat Space-Time,''
JHEP \textbf{10}, 104 (2016)
doi:10.1007/JHEP10(2016)104
[arXiv:1511.01387 [hep-th]].

\bibitem{Bagchi:2016geg}
A.~Bagchi, M.~Gary and Zodinmawia,
``Bondi-Metzner-Sachs bootstrap,''
Phys. Rev. D \textbf{96}, no.2, 025007 (2017)
doi:10.1103/PhysRevD.96.025007
[arXiv:1612.01730 [hep-th]].

\bibitem{Bagchi:2021qfe}
A.~Bagchi, S.~Chakrabortty, D.~Grumiller, B.~Radhakrishnan, M.~Riegler and A.~Sinha,
``Non-Lorentzian chaos and cosmological holography,''
Phys. Rev. D \textbf{104}, no.10, L101901 (2021)
doi:10.1103/PhysRevD.104.L101901
[arXiv:2106.07649 [hep-th]].

\bibitem{Bagchi:2016bcd}
A.~Bagchi, R.~Basu, A.~Kakkar and A.~Mehra,
``Flat Holography: Aspects of the dual field theory,''
JHEP \textbf{12}, 147 (2016)
doi:10.1007/JHEP12(2016)147
[arXiv:1609.06203 [hep-th]].

\bibitem{Ciambelli:2018wre}
L.~Ciambelli, C.~Marteau, A.~C.~Petkou, P.~M.~Petropoulos and K.~Siampos,
``Flat holography and Carrollian fluids,''
JHEP \textbf{07}, 165 (2018)
doi:10.1007/JHEP07(2018)165
[arXiv:1802.06809 [hep-th]].

\bibitem{Bagchi:2019xfx}
A.~Bagchi, A.~Mehra and P.~Nandi,
``Field Theories with Conformal Carrollian Symmetry,''
JHEP \textbf{05}, 108 (2019)
doi:10.1007/JHEP05(2019)108
[arXiv:1901.10147 [hep-th]].






\bibitem{Donnay:2022aba}
L.~Donnay, A.~Fiorucci, Y.~Herfray and R.~Ruzziconi,
``A Carrollian Perspective on Celestial Holography,''
[arXiv:2202.04702 [hep-th]].



\bibitem{Chen:2021xkw}
B.~Chen, R.~Liu and Y.~f.~Zheng,
``On Higher-dimensional Carrollian and Galilean Conformal Field Theories,''
[arXiv:2112.10514 [hep-th]].


\bibitem{Bagchi:2015nca}
A.~Bagchi, S.~Chakrabortty and P.~Parekh,
``Tensionless Strings from Worldsheet Symmetries,''
JHEP \textbf{01}, 158 (2016)
doi:10.1007/JHEP01(2016)158
[arXiv:1507.04361 [hep-th]].

\bibitem{Giddings:1999jq}
S.~B.~Giddings,
``Flat space scattering and bulk locality in the AdS / CFT correspondence,''
Phys. Rev. D \textbf{61}, 106008 (2000)
doi:10.1103/PhysRevD.61.106008
[arXiv:hep-th/9907129 [hep-th]].

\bibitem{Gary:2009mi}
M.~Gary and S.~B.~Giddings,
``The Flat space S-matrix from the AdS/CFT correspondence?,''
Phys. Rev. D \textbf{80}, 046008 (2009)
doi:10.1103/PhysRevD.80.046008
[arXiv:0904.3544 [hep-th]].

\bibitem{Raju:2012zr}
S.~Raju,
``New Recursion Relations and a Flat Space Limit for AdS/CFT Correlators,''
Phys. Rev. D \textbf{85}, 126009 (2012)
doi:10.1103/PhysRevD.85.126009
[arXiv:1201.6449 [hep-th]].

\bibitem{Bagchi:2020fpr}
A.~Bagchi, A.~Banerjee, S.~Chakrabortty, S.~Dutta and P.~Parekh,
``A tale of three \textemdash{} tensionless strings and vacuum structure,''
JHEP \textbf{04}, 061 (2020)
doi:10.1007/JHEP04(2020)061
[arXiv:2001.00354 [hep-th]].

\bibitem{Bagchi:2021rfw}
A.~Bagchi, M.~Mandlik and P.~Sharma,
``Tensionless tales: vacua and critical dimensions,''
JHEP \textbf{08}, 054 (2021)
doi:10.1007/JHEP08(2021)054
[arXiv:2105.09682 [hep-th]].

\bibitem{Sud}
Sudipta~Dutta, {\em{Unpublished notes, 2019.}}

\bibitem{Hao:2021urq}
P.~x.~Hao, W.~Song, X.~Xie and Y.~Zhong,
``A BMS-invariant free scalar model,''
[arXiv:2111.04701 [hep-th]].



\end{thebibliography}
\end{document}